\documentstyle [prl,aps,epsfig,floats]{revtex}

\makeatletter           % Switch to floating figures
\@floatstrue
\def\figure{\let\@capwidth\columnwidth\@float{figure}}
\let\endfigure\end@float
\@namedef{figure*}{\let\@capwidth\textwidth\@dblfloat{figure}}
\@namedef{endfigure*}{\end@dblfloat}
\def\table{\let\@capwidth\columnwidth\@float{table}}
\let\endtable\end@float
\@namedef{table*}{\let\@capwidth\textwidth\@dblfloat{table}}
\@namedef{endtable*}{\end@dblfloat}
\makeatother

\def\ascat{a_{\rm sc}}
\def\alat{a_{\rm lat}}

\def\Tc{T_{\rm c}}
\def\kB{k_{\rm B}}
\def\etal{{\it et al.}}
\def\eff{{\rm eff}}
\def\grad{\mbox{\boldmath$\nabla$}}
\def\x{{\bf x}}
\def\y{{\bf y}}
\def\p{{\bf p}}
\def\U{{\rm U}}
\def\BZ{{\rm BZ}}
\def\lat{{\rm lat}}
\def\phiphic{\langle\phi^2\rangle_{\rm c}}

\begin {document}

%%%%%%%%%%%%%%%%%%%
\def\OurTitlePage
{
\preprint {UW/PT-01-06}

\title {Transition temperature of a dilute homogeneous imperfect Bose
  gas}

\author {Peter Arnold}

\address
    {%
    Department of Physics,
    University of Virginia,
    P.O. Box 400714,
    Charlottesville, VA 22904--4714
    }%

\author {Guy Moore}
\address
    {%
    Department of Physics,
    University of Washington,
    Seattle, Washington 98195--1560
    }%

\date {\today}
\maketitle

\begin {abstract}%
{%
The leading-order effect of interactions on a homogeneous Bose gas is
theoretically
predicted to shift the critical temperature by
an amount $\Delta\Tc \simeq \# \ascat n^{1/3} T_0$ from the
ideal gas result $T_0$, where $\ascat$ is the scattering length and $n$
is the density.
There have been several different theoretical estimates for
the numerical coefficient $\#$.
We claim to settle the issue by measuring the numerical coefficient in a
lattice simulation of O(2) $\phi^4$ field theory in three dimensions---an
effective theory which, as observed previously in the literature, can
be systematically matched to the dilute Bose gas problem to reproduce
non-universal quantities such as the critical temperature.
We find $\#=1.32\pm0.02$.
}%

\end {abstract}
}
%%%%%%%%%%%

\draft
\ifpreprintsty
  \OurTitlePage
  \newpage
\else
  \twocolumn[\hsize\textwidth\columnwidth\hsize\csname
    @twocolumnfalse\endcsname
    \OurTitlePage
  \vskip2pc]
\fi

\begin {figure}
\vbox{
   \begin {center}
      \epsfig{bbllx=224,bblly=107,bburx=525,bbury=609,
              file=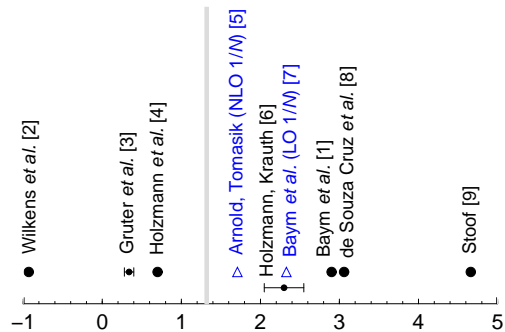,scale=.37,angle=-90}
   \end {center}
   %\vspace*{-.1in}
   \caption{
       Estimates from the literature of the constant $c$ in
       $\Delta\Tc/T_0$ $\to$ $c a n^{1/3}$.  The grey bar is
       the result of this paper.
       \label{fig:literature}
   }
}
\end {figure}

The phase transition temperature $T_0$ of an ideal three-dimensional
Bose-Einstein gas at
fixed density is something that every physicist learns to calculate
in graduate school, if not before.  It is amusing that the first correction
to that result, from arbitrarily weak interactions,
is sufficiently challenging that there has not yet been
theoretical agreement on its magnitude.
It is understood that, in the weak interaction limit, the correction
$\Delta\Tc \equiv \Tc-T_0$ behaves parametrically as
\begin {equation}
   {\Delta\Tc\over T_0} \to c \, \ascat \, n^{1/3} ,
\end {equation}
where $c$ is a numerical coefficient.
(A clean argument may be found in Ref.\ \cite{baym1}.)
Here $\ascat$ is the scattering length, and the weak interaction (or dilute)
limit is $\ascat n^{1/3} \ll 1$: that is, $\ascat$
is small compared to the
typical separation between particles.  (We will assume that the interaction
is repulsive.)
However, there has been little agreement in estimates of
the coefficient $c$,
a variety of which
\cite{baym1,wilkens,gruter,holzmann,arnold,krauth,baymN,SouzaCruz,stoof}
are shown in Fig.\ \ref{fig:literature}.
Some of these estimates are advertised as rough, but others
are not.
%%-0.93 (Wilkens \etal\ \cite{wilkens}),
%%0.34 (Gr\"uter, \etal\ \cite{gruter}),
%%0.7 (Holzmann \etal\ \cite{holzmann}),
%%1.71 (Arnold and Tom\'a\v{s}ik \cite{arnold}),
%%$2.3 \pm 0.25$ (Holzmann and Krauth \cite{krauth}),
%%2.33 (Baym \etal\ \cite {baymN}),
%%2.9 (Baym \etal\ \cite{baym1}),
%%3.059 (de Souza Cruz \etal\ \cite{SouzaCruz}),
%%and
%%4.66 (Stoof \cite{stoof}).
The difficulty
arises
because the phase transition is second order, and
perturbation
theory typically breaks down at second-order phase transitions: the
physics that determines $\Delta\Tc$ is non-perturbative.
As we shall briefly review, the problem of finding $\Delta\Tc$ in the weak
interaction limit can be related
to solving static three-dimensional O(2) scalar $\phi^4$ field theory
\cite{baym1}.
In this paper, we present results from using standard, numerical, lattice Monte
Carlo methods to solve that theory.
In principle this provides an exact method for computing $c$ to any desired
precision; in practice, one is limited by computer time and memory.
Working on desktop computers, we find
$c = 1.32 \pm 0.02$, which is the grey bar in Fig.\ \ref{fig:literature}.

It is long distance physics that determines $\Delta\Tc$ in the weak
interaction limit.
It is well known \cite {review}
that, at distance scales large compared to the scattering
length $\ascat$, an appropriate effective theory for a dilute Bose gas is the
second-quantized Schr\"odinger equation, together with a chemical potential
$\mu$ that couples to particle number density $\psi^*\psi$, and a
$|\psi|^4$ contact interaction that reproduces low-energy scattering.
The corresponding Lagrangian is
\begin {equation}
   {\cal L} = \psi^* \left(
        {i\hbar} \, \partial_t + {\hbar^2\over 2m} \, \nabla^2
        + \mu \right) \psi
    - {2\pi\hbar^2 \ascat\over m} \, (\psi^* \psi)^2 .
\label {eq:L1}
\end {equation}
Corrections to this effective theory (due, for instance, to energy
dependence of the cross-section or 3-body
interactions)
may be ignored for the propose of
computing the leading-order result for $\Delta\Tc$.
To study (\ref{eq:L1}) at finite temperature, apply the imaginary
time formalism, so that $t$ becomes $i\tau$ and imaginary time $\tau$
is periodic with period $\hbar\beta = \hbar/\kB T$.
The field $\psi$ can then be decomposed into frequency modes with Matsubara
frequencies $\omega_n = 2\pi n/\hbar\beta$.
For distances large compared to the thermal wavelength
$\lambda = \hbar \sqrt{2\pi\beta/m}$, and sufficiently near the
transition so that $|\mu| \ll T$, the non-zero Matsubara frequencies
decouple from the dynamics, leaving behind an effective theory of
only the zero-frequency modes $\psi_0$, with the action
$
   S =
   \hbar^{-1} \int_0^{\hbar\beta} d\tau \int d^3x \> {\cal L}
$
becoming \cite{baym1}
\begin {equation}
   \beta \int d^3x \> \left[ \psi_0^* \left(
        - {\hbar^2\over 2m} \, \nabla^2
        - \mu_\eff \right) \psi_0
    + {2\pi\hbar^2 \ascat\over m} \, (\psi_0^* \psi_0)^2 \right] ,
\label {eq:L2}
\end {equation}
up to corrections that again do not affect the leading-order result
for $\Delta\Tc$.
This action can be thought of as the $\beta H$ of a classical
three-dimensional field theory.
Finally, it is convenient to rewrite
$\psi_0 = (\phi_1+i\phi_2) \sqrt{2\pi}/\lambda$ so that
the effective action becomes a conventionally normalized
O(2) field theory:
\begin {equation}
   S = \int d^3x \> \left[{1\over2} \, |\grad\phi|^2 + {1\over2} \, r \phi^2
             + {u\over 4!} \, (\phi^2)^2 \right] ,
\label{eq:O2}
\end {equation}
where $\phi$ is understood to be a 2-component real
vector $(\phi_1,\phi_2)$ and $u = 96 \pi^2 \ascat /\lambda^2$.

As noted by Baym \etal \cite{baym1}, it is technically somewhat more
convenient, in this formalism, to
calculate the shift $\Delta n_{\rm c}(T)$ in the critical density
at fixed temperature instead of the shift
$\Delta\Tc(n)$ in the critical temperature at fixed density.
The two are trivially related at first order in $\Delta\Tc$ by
$\Delta\Tc/T = -{2\over3}\Delta n_{\rm c}/n$, where the factor of
${2\over3}$ arises from the ideal gas relation $T_0 \propto n^{2/3}$.
In the field theory, $n$ is given by $\langle \psi^*\psi\rangle$,
which is proportional to $\langle \phi^2 \rangle$.  Putting everything
together \cite{baym1},
\begin {equation}
   {\Delta \Tc\over T_0} = - {2 m \kB T_0 \over 3\hbar^2 n} \,
        \Delta\phiphic ,
\label {eq:dTc}
\end {equation}
where
\begin {equation}
   \Delta\phiphic \equiv \left[\phiphic\right]_u - \left[\phiphic\right]_0
\end {equation}
%represents 
is
the difference between
the effective theory value of $\langle\phi^2\rangle$, at the critical
point, for the cases of (i) $u$ small and (ii) the ideal gas $u{=}0$.
Unlike $\phiphic$,
the difference $\Delta\phiphic$ is an infrared quantity,
independent of how the effective theory (\ref{eq:O2}) is regularized in
the ultraviolet (UV).

Finding $\left[\phiphic\right]_u$ in the effective theory (\ref{eq:O2})
corresponds to fixing $u$, varying $r$ to reach the
critical point, and then measuring $\langle\phi^2\rangle$.  The only
parameter of this problem is $u$.  By dimensional analysis,
physics is therefore non-perturbative at the infrared length scale
$1/u$, and, again by dimensional analysis,
$\Delta\phiphic$ is proportional to $u$.
Putting this together with the ideal gas formula
$T_0 = (2\pi\hbar^2/\kB m) [n/\zeta\!\left(3\over2\right)]^{2/3}$, one may
summarize the relationship between the weak interaction limit for
$\Delta\Tc$ and the O(2) effective theory as
$\Delta\Tc/T_0 \to c\,\ascat\,n^{1/3}$ with
\begin {equation}
   c = - {128 \pi^3 \over \left[\zeta({3\over2})\right]^{4/3}}
         \, {\Delta\phiphic \over u} \,.
\label{eq:c}
\end {equation}

To compute $\Delta\phiphic/u$, we
put the O(2) theory (\ref{eq:O2}) on a lattice.  For example, the
most straightforward discretization would use the action
\begin {eqnarray}
   S_\U = \alat^3 \sum_{\langle\x\y\rangle} {1\over2} \left[
         \left(\phi_{1\x}-\phi_{1\y}\over \alat\right)^2
         + \left(\phi_{2\x}-\phi_{2\y}\over \alat\right)^2
       \right]
\nonumber\\
       + \alat^3 \sum_\x \left[
          {r_\lat\over 2} \, (\phi_{1\x}^2+\phi_{2\x}^2)
          + {u\over4!} \, (\phi_{1\x}^2+\phi_{2\x}^2)^2
       \right] ,
\label {eq:SlatU}
\end {eqnarray}
on a simple cubic lattice, where $\langle\x\y\rangle$ represents all
nearest-neighbor pairs and $\alat$ is the lattice spacing
(unrelated to $\ascat$).
The dimensionless coupling of the lattice theory is $u\alat$, and the
continuum limit is $u\alat \to 0$.

Our simulations use an improved action to reduce lattice spacing
errors.
(The subscript U on $S$ in Eq.\ (\ref{eq:SlatU}) stands for ``unimproved.'')
Details concerning the action and our simulations are given in
Ref.\ \cite{boselat2}.
Our simulations use a combination of heat bath and multi-grid updates
\cite{lattice}.
At finite volume, we use the method of Binder cumulants
\cite{boselat2,BinderCumulants}
to determine a nominal critical value for $r$.

Because three-dimensional scalar theory requires UV regularization
of its $\phi^2$ interactions, the $r_\lat$ in the lattice action
(\ref{eq:SlatU}) is not simply the $r_\eff$ in the effective theory
action (\ref{eq:O2}), which in turn is not simply related to the
chemical potential $\mu$ in the original action (\ref{eq:L1}).
%However, we needn't bother ourselves here with deriving the actual relations
%between these quantities \cite{foot:r}
%PETER:  the following is redundant and not clear enough even after changes.
%%However, this relation is irrelevant for our purposes here
%%\cite{foot:r}, because
%%we simply wish to adjust the chemical
%%potential so that we are at the critical temperature, where we will
%%then measure $\Delta\phiphic$.  So, in a lattice theory
%%such as (\ref{eq:SlatU}), we need not care about the exact relation
%%between $r_\lat$ and $\mu$, but we simply wish to adjust $r_\lat$ for
%%fixed $u$ until we find the critical point, and then measure
%%$\Delta\phiphic$.
However, for our purposes here, we are only interested in adjusting
$r_\lat$ to find the critical point, for a given $u$, and measuring
$\Delta\langle\phi^2\rangle$ there.  The actual relation between
$r_\lat$, $r_\eff$, and $\mu$ is unnecessary
\cite{foot:r}.
The $\phi^4$ interactions, in contrast, do not
require UV regularization: in the limit $u\alat \to 0$, the coupling
$u$ of the lattice theory may be identified with the continuum coupling
$u=96\pi^2\ascat/\lambda^2$ introduced earlier.

For a given $u$, we compute $\phiphic$ by Monte Carlo
numerical simulations of the lattice theory.  The $u{=}0$ piece of the
difference $\Delta\phiphic$ can be easily computed without
simulations:
\begin {equation}
   \Delta\phiphic
   = \lim_{\alat\to0} \left[
        \langle\phi^2\rangle_\lat - \int_{\p\in\BZ} G^{ii}_\lat(\p)
   \right] ,
\end {equation}
where $G^{ij}_\lat(\p)$ is the free lattice propagator, and the momentum $\p$
is integrated over the Brillouin zone (BZ).  Such integrals
are reviewed in Ref.\ \cite{boselat2}.  For the unimproved lattice theory
(\ref{eq:SlatU}), for example \cite{watson},
\begin {equation}
   \Delta\phiphic
   = \lim_{\alat\to0} \left[
        \langle\phi^2\rangle_\lat - {\Sigma_\U\over2\pi \alat} 
   \right],
\end {equation}
where $\Sigma_\U = 3.175~911~535~625~\cdots$.

\begin {figure}
\vbox{
   \begin {center}
      \epsfig{file=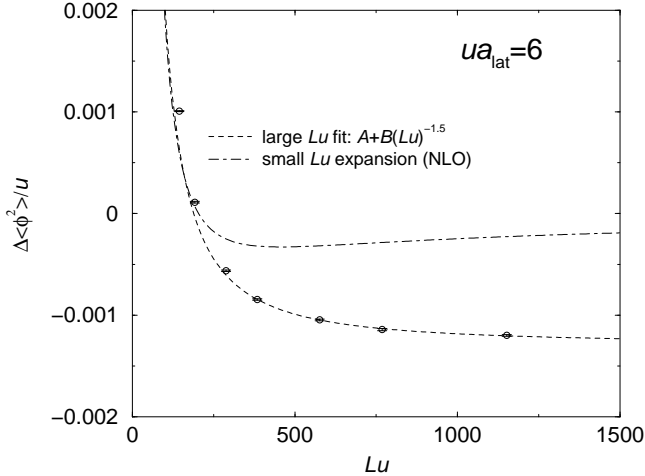,scale=.38,angle=-90}
   \end {center}
   \vspace*{-.1in}
   \caption{
       $O(u\alat)$-corrected results for $\Delta\phiphic$ vs.\ system size
       at $u\alat=6$.
       A numerical fit to large $L$ scaling behavior is shown, which fits
       the last 4 points with confidence level 61\%.
       Also shown, for comparison, is
       a small $Lu$ expansion \protect\cite{boselat2}
       of the exact continuum result in finite volume
       (at next-to-leading order in $Lu$).
       \label{fig:ua6}
   }
}
\end {figure}

\begin {figure}
\vbox{
   \begin {center}
      \epsfig{file=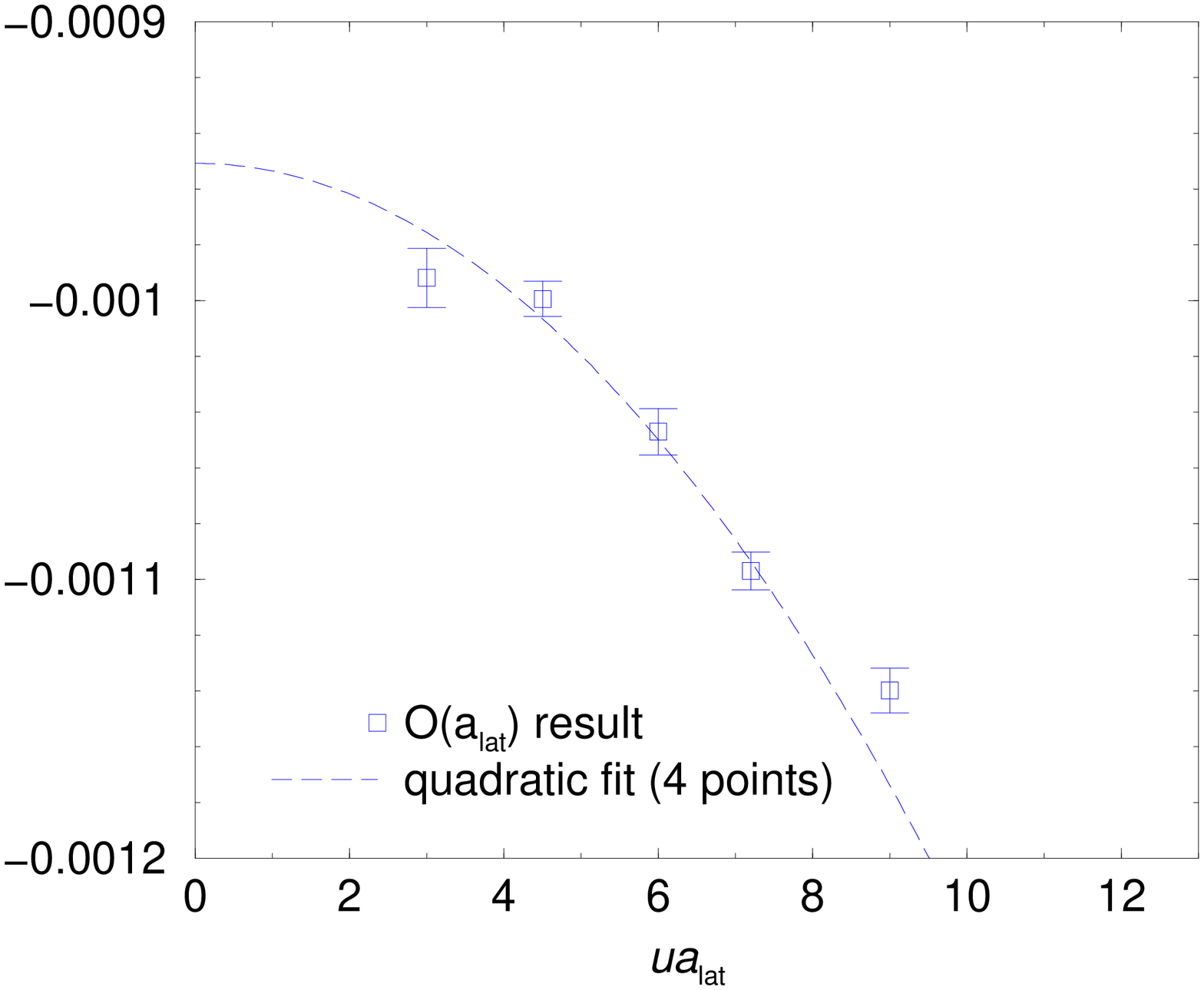,scale=.40,angle=-90}
   \end {center}
   %\vspace*{-.1in}
   \caption{
       Results for $\Delta\phiphic$ vs.\ $u\alat$ at $Lu=576$ that
       incorporate perturbatively calculated $O(u\alat)$ corrections.
       The line is a fit of
       $A + B(u\alat)^2$ to all but the rightmost data point
       and has confidence level 14\%.
       \label{fig:Lu576}
   }
}
\end {figure}

There are two limits that must be taken of lattice Monte Carlo data:
the continuum limit $u\alat{\to}0$ and the infinite volume limit
$Lu{\to}\infty$.
Fig.\ \ref{fig:ua6} shows the dependence of our data on system size
($Lu$) at $u\alat=6$.
Fig.\ \ref{fig:Lu576} shows the dependence on lattice spacing
($u\alat$) at $Lu=576$.
From these two figures, it is reasonably clear that our raw data
includes reasonably large volumes and reasonably small lattice spacings.
We will discuss extrapolations of the infinite-volume continuum limit
using finite volume scaling, the known critical exponents of this model, and
an analysis of finite lattice-spacing errors.

\begin {table}
  \begin {center}
  \setlength{\tabcolsep}{10pt}
  
  \begin {tabular}{|c|l|}                          \hline
  ($Lu$, $u\alat$) & \multicolumn{1}{c|}{$\Delta\phiphic/u$} \\ \hline
  (576, 6)  & $-0.001047(8)$ \\
  (576, 3)  & $-0.000992(11)$ \\
  (1152, 6) & $-0.001200(9)$ \\       \hline
  \end {tabular}
  \end {center}
  \caption{
     Selected data for $\Delta\phiphic/u$.
     \label{tab:selected}
  }
\end {table}

First, however, we wish to show that one can obtain a quick
estimate from the raw data without relying on anything fancy.
Table \ref{tab:selected} shows values associated with a selected subset
of the largest $Lu$ and smallest $u\alat$ data points from the figures.
We take the $(Lu,u\alat)=(576,6)$ result as a starting point for our
estimate.  The finite-volume correction is at least as big as the
difference with the value at $(1152,6)$ [which corresponds to
doubling $Lu$] but is unlikely to be double this difference.
This difference is roughly $-0.00015$ (ignoring the small statistical errors),
so we might estimate the finite-volume correction to the $Lu=576$ value
to be somewhere between $-0.00015$ and $-0.00030$.
From a similar comparison of the (576,6) and (576,3) data, we might
estimate the finite lattice spacing correction to (576,6) to be
between roughly $+0.00005$ and $+0.00011$.
Adding our corrections and the original $(576,6)$
data point, the final rough estimate of the
continuum infinite-volume value would be
\begin {equation}
   {\Delta\phiphic\over u} \simeq -0.00119 \pm 0.00011 ,
\end {equation}
which, by (\ref{eq:c}) would translate to $c = 1.31 \pm 0.12$.

\begin {figure}
\vbox{
   \begin {center}
      \epsfig{file=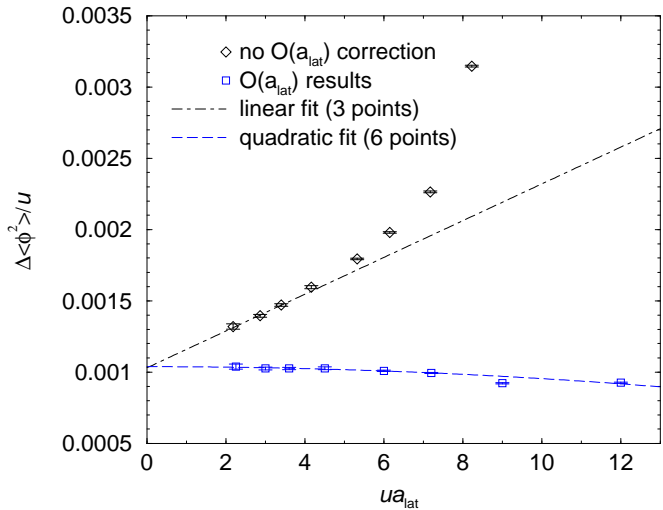,scale=.40,angle=-90}
   \end {center}
   %\vspace*{-.1in}
   \caption{
       The squares show $\Delta\phiphic$ vs.\ $u\alat$ at
       $Lu=144$
       incorporating $O(u\alat)$ corrections.
       The line through them is a fit of the first 6 points to
       $A + B(u\alat)^2$ and has confidence level 94\%.
       The diamonds represent the corresponding uncorrected data, with
       a straight line fit to the first 3 points to guide the eye.
       \label{fig:Lu144}
   }
}
\end {figure}

We now summarize a more careful analysis of corrections,
detailed in ref.\ \cite{boselat2}.
Our strategy is to start again with $Lu=576$ data,
extrapolate the continuum limit, and
estimate the finite volume correction.
To improve the approach to the continuum limit, we have
analytically calculated the $O(u\alat)$ corrections to
$\Delta\phiphic$
and the relation between lattice and continuum values of $u$,
using lattice perturbation theory \cite{boselat2}.
Fig.\ \ref{fig:Lu144} shows $Lu=144$ data which clearly demonstrates
our control of lattice spacing errors.  The uncorrected data clearly
has a linear dependence on $u\alat$.
But the corrected data fits, to high confidence level, the assertion
that the remaining error scales as $(u\alat)^2$.
Based on a similar fit to the data of Fig.\ \ref{fig:Lu576},
we estimate the $u\alat\to0$ result at $Lu=576$ as
\begin {equation}
   \left[\Delta\phiphic\over u\right]_{Lu=576} = -0.000957\pm0.000015 .
\label {eq:Lu576}
\end {equation}

Finite scaling arguments predict that the large $Lu$ corrections to
$\Delta\phiphic$ should scale as \cite{boselat2}
\begin {equation}
   \Delta\phiphic \sim A + B  L^{-(1-\alpha)/\nu} 
%        + L^{-1/\nu-\omega} \left(C + D L^{\alpha/\nu}\right)
\end {equation}
if the method of Binder cumulants is used to determine the transition point
in finite volume.
Here
$\alpha \simeq -0.01$
and $\nu = (2-\alpha)/3 \simeq 0.67$ are the
specific heat and correlation length
critical exponents of the
O(2) model \cite{exponents}.
Further discussions of fits, and an analysis of corrections to scaling,
may be found in ref.\ \cite{boselat2}.  On the basis of these fits,
we estimate the finite size correction at $Lu=576$ to be
$0.000241 \pm 0.000007$ .
Putting this together with the $Lu=576$ continuum extrapolation
(\ref{eq:Lu576}), we obtain the final infinite-volume continuum result
\begin {equation}
   {\Delta\phiphic\over u} = -0.001198\pm0.000017 .
\end {equation}
Using Eq.\ (\ref{eq:c}) for the weak-interaction limit of $\Delta\Tc$,
the result for $\Delta\phiphic$ translates to
\begin {equation}
   {\Delta\Tc\over T_0} \to (1.32 \pm 0.02) \, \ascat n^{1/3} .
\end {equation}

It is interesting to compare our numerical results with results from
the large $N$ expansion, depicted by triangles in Fig.\ \ref{fig:literature}.
In large $N$, one generalizes the O(2) effective
field theory to an O($N$) theory of $N$ real fields,
calculates results in powers of $1/N$, and hopes
the expansion will be useful for the case of interest, $N=2$.
The leading-order (LO) result $c \simeq 2.33$ \cite {baymN}
is off by 77\%, but the next-to-leading-order
(NLO) result $c \simeq 1.71$ \cite{arnold}
moves in the right direction and is off by only 30\%.  This is surprisingly
good for an expansion that treats $N=2$ as large.

We should comment on the discrepancy of our results with
previous numerical calculations in the literature \cite{gruter,krauth},
shown in Fig.\ \ref{fig:literature},
which used a radically different starting point.
Rather than using field theory methods and the grand canonical ensemble,
they start with the path integral for a large, fixed
number of particles in a box.
The recent work of Holzmann and Krauth \cite{krauth},
however, makes a flawed assumption
at the very beginning: they assume that the integrand of the path integral
can be expanded perturbatively in the interaction%
%.  They keep only the
%leading term, and so automatically obtain a result for $\Delta\Tc$
%proportional to $\ascat$.  
, and keep only the leading term.
This is wrong because interactions cannot,
generically, be treated perturbatively at a second-order phase transition.

We believe one likely problem with the older simulations of
Gr\"uter \etal\ is inadequate system size \cite{boselat2}.
Reppy \etal \cite{reppy} have reported
$c=5.1\pm0.9$ from experimental data
on He-Vycor systems, but cautions
about the data's interpretation may be found in Ref.\ \cite{arnold}.

As we completed this work, another paper appeared \cite{verynew} which
uses techniques very similar to ours
and obtains the statistically compatible result
$c=1.29\pm0.05$.

This work was supported by the U.S. Department
of Energy under Grant Nos.\ DE-FG03-96ER40956
and DE-FG02-97ER41027.


\begin {references}

\bibitem{baym1}
   G.\ Baym, J.-P. Blaizot, M. Holzmann, F. Lalo\"e, and D. Vautherin, 
   Phys.\ Rev.\ Lett.\ {\bf 83}, 1703 (1999).

\bibitem {wilkens}
  M. Wilkens, F. Illuminati, and M. Kr\"amer,
  J. Phys.\ B {\bf 33}, L779 (2000).

\bibitem {gruter}
  P. Gr\"uter, D. Ceperley, and F. Lalo\"e,
  Phys.\ Rev.\ Lett.\ {\bf 79}, 3549 (1997).

\bibitem {holzmann}
  M. Holzmann, P. Gr\"uter, and F. Lalo\"e,
  Euro.\ Phys.\ J. B {\bf 10}, 739 (1999).

\bibitem {arnold}
  P. Arnold and B. Tom\'a\v{s}ik,
  Phys.\ Rev.\ {\bf A62}, 063604 (2000).

\bibitem {krauth}
   M. Holzmann and W. Krauth,
   Phys.\ Rev.\ Lett. {\bf 83}, 2687 (1999).

\bibitem {baymN}
   G.\ Baym, J.-P.\ Blaizot, and J.\ Zinn-Justin, 
   Europhys.\ Lett.\ {\bf 49}, 150 (2000).

\bibitem{SouzaCruz}
   F. de Souza Cruz, M. Pinto, and R. Ramos,
   cond-mat/0007151 (unpublished).

\bibitem{stoof}
   H.T.C.\ Stoof, Phys.\ Rev.\ A {\bf 45}, 8398 (1992);
   M.\ Bijlsma and H.T.C.\ Stoof, Phys.\ Rev.\ A {\bf 54}, 5085 (1996).

\bibitem{review}
   F. Dalfovo, S. Giorgine, L. Pitaevski, and S. Stringari,
   Rev.\ Mod.\ Phys.\ {\bf 71}, 463 (1999).

\bibitem{boselat2}
   P. Arnold and G. Moore, cond-mat/0103227.

\bibitem{lattice}
J.~Goodman and A.~D.~Sokal,
%``Multigrid Monte Carlo Method For Lattice Field Theories,''
Phys.\ Rev.\ Lett.\ {\bf 56}, 1015 (1986);
%%CITATION = PRLTA,56,1015;%%
J.~Goodman and A.~D.~Sokal,
%``Multigrid Monte Carlo Method: Conceptual Foundations,''
Phys.\ Rev.\ D {\bf 40}, 2035 (1989).
%%CITATION = PHRVA,D40,2035;%%

\bibitem{BinderCumulants}
   K. Binder,
   Phys.\ Rev.\ Lett.\ {\bf 47}, 693 (1981).

\bibitem{foot:r}
   The relationship between $r_\lat$ and $r_\eff$ is derived in
   Ref.\ \cite{boselat2}, which uses this relation to give a lattice simulation
   result for $r_\eff$.  The relationship between $r_\eff$ and $\mu$ is
   derived in Ref.\ \cite{trap}, which gives a result for $\mu$ at the
   transition and uses this to calculate the second-order correction to
   $\Tc$ for a Bose gas in a harmonic trapping potential.

\bibitem{trap}
   P. Arnold and B. Tom\'a\v{s}ik,
   cond-mat/0105147.


\bibitem{watson}
   G. Watson,
   Quart.\ J.\ Math.\ (Oxford, 1st series) {\bf 10}, 266 (1939).

\bibitem{exponents}
   R. Guida and J. Zinn-Justin,
   J.\ Phys.\ A: Math. Gen.\ {\bf 31}, 8103 (1998).

\bibitem{reppy}
   J. Reppy, B. Crooker, B. Hebral, A. Corwin, J. He, and G. Zassenhaus,
   Phys.\ Rev.\ Lett.\ {\bf 84}, 2060 (2000).

\bibitem{verynew}
   V.A. Kashurnikov, N. Prokof'ev, and B. Svistunov, cond-mat/0103149;
   N. Prokof'ev, and B. Svistunov, cond-mat/0103146.


\end {references}
\end {document}